\documentclass[twocolumn,superscriptaddress]{revtex4-2}      
\usepackage{amsmath,mathtools,amsthm}
\usepackage{amssymb}
\usepackage[mathscr]{eucal}
\usepackage{bm,booktabs}
\usepackage{bbm}
\usepackage[shortlabels]{enumitem}
\usepackage{hyperref}
\usepackage[usenames,dvipsnames]{xcolor}

\usepackage{tikz}
\usepackage[caption=false]{subfig}
\usepackage{graphicx}
\usepackage{times}

\newcommand{\me}{\textup{e}}

\begin{document}
\title{Network processes on clique-networks with high average degree: the limited effect of higher-order structure}
\author{Clara Stegehuis}
\affiliation{Department of Electrical Engineering, Mathematics and Computer Science, Twente University, The Netherlands}
\author{Thomas Peron}
\affiliation{Institute of Mathematics and Computer Science, University of S\~ao Paulo,
S\~ao Carlos 13566-590, S\~ao Paulo, Brazil}
\date{\today}

\begin{abstract}
    In this paper, we investigate the effect of local structures on network processes. We investigate a random graph model that incorporates local clique structures to deviate from the locally tree-like behavior of most standard random graph models. For the process of bond percolation, we derive analytical approximations for large outbreaks and the critical percolation value. Interestingly, these derivations show that when the average degree of a vertex is large, the influence of the deviations from the locally tree-like structure is small. Our simulations show that this insensitivity to local clique structures often already kicks in for networks with average degrees as low as 6. Furthermore, we show that the different behavior of bond percolation on clustered networks compared to tree-like networks that was found in previous works can be almost completely attributed to differences in degree sequences rather than differences in clustering structures. We finally show that these results also extend to completely different types of dynamics, by deriving similar conclusions and simulations for the Kuramoto model on the same types of clustered and non-clustered networks. 
\end{abstract}

\maketitle

\section{Introduction}

One of the problems that has motivated research in network science to a large extent is the assessment of how structural characteristics of real-world networks determine the performance of dynamical processes that take place on them~\cite{newman2018networks}. Most analytical approaches to this problem use networks constructed via configuration models~\cite{bollobas1980probabilistic} as the substrate for the dynamics. In such models, one specifies the fraction of vertices with $k$ neighbors $p_k$. A sequence of vertex degrees $\{k_1,...,k_N\}$ is then drawn independently following $p_k$. The network is then assembled by choosing pairs of ``half-edges'' (or stubs) uniformly at random from this sequence, which are joined to form complete edges. While this method is able to generate networks with any prescribed degree distribution along with offering great analytic tractability, it has the shortcoming that the generated networks are locally tree-like. That is, the density of cycles vanishes asymptotically as the network size increases. This contrasts markedly with the rich topological structure of real-world networks, which often exhibit short cycles, degree correlations and clustering (i.e., the tendency of groups of three vertices to form triangles). Clustering is common to a variety of systems, but it is specially important in social networks, where the average probability that two neighbors of a vertex are also neighbors themselves (also referred to as clustering coefficient) often reaches values of tens of percent~\cite{newman2018networks}. Other classes of systems known to be highly clustered in this sense comprise biological and information networks~\cite{newman2018networks}. Hence, the inclusion of triangles and other types of subgraphs in random network models appears to be a crucial step to model dynamical process on networks accurately.

A practical method to create analytically tractable random networks with a more realistic clustering structure is to extend the standard configuration in order to explicitly include the generation of motifs that yield clustering. The first model of this kind was 
proposed independently by Newman~\cite{newman2009} and Miller~\cite{miller2009percolation}. This model sets two degree sequences drawn from a joint degree distribution: the first sequence prescribes how many edges each vertex is incident to, exactly as in the standard configuration model; and the second degree sequence defines the number of triangles to which each vertex is attached. As the model then matches these stubs accordingly into edges and triangles, it generates networks with non-vanishing clustering even in the limit of large sizes~\cite{newman2009,miller2009percolation}. This strategy can be adapted to produce networks not only with triangles, but also with distributions of cliques of larger size~\cite{gleeson2009bond}, different types of subgraphs~\cite{karrer2010}, or edge-multiplicities~\cite{zlatic2012networks}.

A number of previous authors have investigated the impact of added clustering on several types of network dynamics by employing such extensions of the standard configuration model. For instance, using a model that created networks with arbitrary distributions of cliques~\cite{gleeson2009bond}, Gleeson et al.~\cite{gleeson2010clustering} showed that clustered networks exhibit higher bond percolation thresholds in comparison to locally tree-like structures with same degree distributions and correlation properties. Very recently, Mann et al.~\cite{mann2021random} studied the percolation properties of the model by Karrer and Newman~\cite{karrer2010} under different combinations of cycles and cliques as building blocks for the networks. The authors confirmed that the increased clustering created by cliques leads to higher percolation thresholds~\cite{mann2021random}. On the other hand, the dynamics of networks containing only cycles were shown to approach the result obtained for the configuration model when the length of these cycles increases, as the model then becomes more locally tree-like. A different method to add clique structures to standard configuration models is to use household models, where every vertex of the configuration model is exploded into a clique of a specified size~\cite{ball2009}. In this model, clustering was found to increase the percolation threshold~\cite{ball2010,coupechoux2014}. However, when including other clustered subgraphs than cliques, the percolation threshold may either increase or decrease compared to a locally tree-like model~\cite{hofstad2015,stegehuis2016}.

Network processes on configuration models with higher-order clustering find their widest application in mathematical epidemiology, 
 because of the natural importance of modeling of outbreaks in real-world scenarios and the close analogy between disease spreading and percolation processes.
 Indeed, many results uncovered in the context of percolation have counterparts in disease spreading. For instance, the presence of triangles has been found to increase the epidemic threshold while decreasing the outbreak size
~\cite{miller2009percolation}. Likewise, networks composed of cycles have been shown to yield epidemic dynamics similar to those of tree-like networks as the length of these cycles increases~\cite{ritchie2016beyond}. Examples of other dynamics investigated with higher-order configuration models include cascade propagation
~\cite{hackett2011cascades,hackett2013cascades}, the Ising model~\cite{herrero2015ising}, and synchronization of coupled oscillators~\cite{peron2013synchronization}. 

In this paper we reveal an effect that seems to have remained unnoticed in previous works; namely, we show that the influence of higher-order subgraphs on network dynamics is negligible when the average degree is large. Specifically, we show that in such a limit the percolation dynamics of clustered networks for large outbreaks as well as the critical percolation value converge to the one expected for locally tree-like networks. We focus on the most clustered subgraphs possible: cliques of different sizes. While our analytical results are for the large average degree limit, our simulations show that this convergence kicks in for average degrees as small as 6 for several degree distributions. We also show that these conclusions hold for the synchronization transition of phase oscillators modelled by the Kuramoto model~\cite{rodrigues2016kuramoto}, indicating that the insensitivity to local network structures may hold for a wide range of network processes.    

\paragraph*{Organization of the paper.}
We first describe the random graph model with subgraphs in Section~\ref{sec:model}. We then focus on the setting where the network is formed by $k$-cliques of one given size. In Section~\ref{sec:perccliques}, we show that in such networks, size of the largest component under percolation becomes \emph{independent} of the clique structures under large outbreaks. We then turn to small outbreaks in Section~\ref{sec:piccliques}, where we show that the critical percolation threshold also can be approximated by a $k$-independent value when the average degree of the network is large. We then investigate a setting where different clique sizes are present, in Section~\ref{sec:mixedcliquecomps}. We show that even in this setting, where it has been reported that the possible introduction of degree correlations can affect the size of the largest component under percolation, when the average degree grows, large outbreaks only depend on the degree distribution of the network, not on the specific clique sizes. Finally, in Section~\ref{sec:Kuramoto}, we use analytical approximations as well as simulations to show that for a very different network process, the Kuramoto model, this insensitivity for local clustered network structures also appears for networks of large average degrees.

\section{Random graph model with clique subgraphs}\label{sec:model}
As a random graph model, we employ the random graph model with clustering developed in~\cite{newman2009,karrer2010}. This random graph model is a general framework that extends the configuration model to create networks with specified densities of arbitrary specified subgraphs. Including clustered subgraphs in the set of specified subgraphs enables to overcome the locally tree-like property of the standard configuration model. 
In this manuscript we focus on the most clustered sets of subgraphs, cliques. That is, every vertex has a joint clique degree vector $(s^{(1)},\dots,s^{(m)})$. Here $s^{(1)}_i$ denotes the edge-degree of vertex $i$, and $s^{(j)}_i$ denotes the clique-degree of size $j+1$ of vertex $i$. The clique-degree of a vertex describes a vertex' involvement in cliques of a specified size. Thus, a vertex of clique degree $s^{(2)}_i=3$ is part of 3 cliques of size 3. We denote the probability that a vertex has clique-degrees $s^{(1)},\dots,s^{(k)}$ by $q_{s^{(1)},\dots,s^{(m)}}$. The degree or the total number of connections of vertex $i$ is then described by $\sum_{j=1}^mjs^{(j)}_i$, because every clique of size $j+1$ adds $j$ connections to the vertex. We denote the degree distribution of a vertex by $p_k$, so that 
\begin{equation}\label{eq:pk}
	p_k=\sum_{s^{(1)},\dots,s^{(k)}=1}^\infty q_{s^{(1)},\dots,s^{(m)}} \mathbbm{1}_{s^{(1)}+2s^{(2)}+\dots+ms^{(m)}=k}.
\end{equation}

After sampling a joint clique-degree for every vertex, the network is then formed by selecting $j$ uniformly chosen clique-edges of size $j$, and pairing the corresponding vertices into a clique for all $j$ until all clique-edges have been paired into a clique. This is an extension of the standard configuration model, where the network is formed by pairing two uniformly chosen half-edges until all half-edges have been paired.

\section{Bond percolation with general cliques}\label{sec:perccliques}
We now investigate the behavior of this network model under bond percolation, where every edge is removed independently with probability $1-\pi$. We first focus on the case where every vertex is part of only $k$-cliques. Let $q_i$ denote the probability that a randomly chosen vertex is part of $i$ $k$-cliques. Define the generating functions
\begin{equation}
	g(x)=\sum_{i=1}^{\infty}q_ix^i, \quad g_p(x)=\frac{1}{\langle s \rangle}\sum_{i=1}^{\infty}iq_ix^{i-1}=\frac{g'(x)}{\langle s \rangle},
\end{equation}
where $\langle s\rangle$ denotes the average number of $k$-cliques a vertex is part of. Let $u$ denote the probability that a randomly chosen clique-edge is not connected to the giant component. We are interested in the fraction of vertices in the largest component after percolation, $S$, which can be obtained by~\cite{newman2009}, 
\begin{align}\label{eq:cliqueuS}
	u&=g_p(\sum_{j=0}^{k-1}h(k,j,\pi)u^j),\nonumber\\
	S&=1-g(\sum_{j=0}^{k-1}h(k,j,\pi)u^j),
\end{align}
where $h(k,j,\pi)$ is the probability that a given vertex of a $k$-clique is still connected to $j$ other vertices of the clique after percolation with probability $\pi$. These implicit equations are in general difficult to solve~\cite{karrer2010,mann2021exact}, so that it is difficult to make general observations on the solution of these equations. Therefore, we here focus on an approximation of $S$, first for large outbreaks ($\pi$ large), and then for small outbreaks (approximating the critical value where $S$ becomes larger than zero). In these approximations, we will assume that the number of connections of a vertex is large.

When the degree of a vertex is large, the probability that a randomly chosen clique-edge is not connected to the giant component becomes small. Thus, we expand~\eqref{eq:cliqueuS} with a first-order Taylor expansion around $u=0$. This yields
\begin{equation}
	u=g_p(h(k,0,\pi)+h(k,1,\pi)u).
\end{equation}
Filling in the expressions $h(k,0,\pi)=(1-\pi)^{k-1}$ and $h(k,1,\pi)=(k-1)\pi(1-\pi)^{2(k-2)}$ yields
\begin{align}
	u&\approx g_p((1-\pi)^{k-1}+u\pi(1-\pi)^{2(k-2)})\nonumber\\
	& \approx g_p((1-\pi)^{k-1})\nonumber\\
	& \quad +g_p'((1-\pi)^{k-1})(k-1)\pi(1-\pi)^{2(k-2)}u
\end{align}
This results in 
\begin{equation}
	u\approx \frac{g_p((1-\pi)^{k-1})}{1-g_p'((1-\pi)^{k-1})(k-1)\pi(1-\pi)^{2(k-2)}}.
\end{equation}
Using a first order Taylor expansion,~\eqref{eq:cliqueuS} then yields for $S$,
\begin{align}\label{eq:Sgeneral}
	S& \approx 1-g(h(k,0,\pi)+h(k,1,\pi)u)\nonumber\\
	& \approx 1-g(h(k,0,\pi)) +g'(h(k,0,\pi))h(k,1,\pi)u\nonumber\\
	& = 1-g((1-\pi)^{k-1})\nonumber\\
	& \quad -\frac{g_p((1-\pi)^{k-1})(k-1)(1-\pi)^{2(k-2)}\pi g'((1-\pi)^{k-1})}{1-g_p'((1-\pi)^{k-1})(k-1)\pi(1-\pi)^{2(k-2)}}\nonumber\\
	& = 1-g((1-\pi)^{k-1})\nonumber\\
	& \quad -\frac{\langle s \rangle g_p((1-\pi)^{k-1})^2(k-1)(1-\pi)^{2(k-2)}\pi }{1- g_p'((1-\pi)^{k-1})(k-1)\pi(1-\pi)^{2(k-2)}},
\end{align}
where $\langle s \rangle$ again denotes the average number of cliques a vertex is part of. 

Now $g((1-\pi)^{k-1})=g_D(1-\pi)$, where $g_D(x)=\sum_kp_kx^k$ is the generating function of the vertex degrees from~\eqref{eq:pk}. This means that for a given degree distribution $D$, the leading order term of the approximation of the largest component size does not depend on the clique size in which the vertex degrees are split. Furthermore, we show in Appendix~\ref{sec:eqsecondterm} that the numerator of the second term also only depends on the degree distribution, not on the clique structure. Furthermore, $g_p'((1-\pi)^{k-1})$ decreases when the network degrees increase. Thus, large outbreaks become asymptotically independent of the clique structures in the networks.

\paragraph{Example: Regular degrees.}
We now apply our approximations to several frequently used degree distributions. In regular networks, every vertex is part of $s$ $k$-cliques. Then, $g(x)=x^s$ and $g_p(x)=x^{s-1}$, so that~\eqref{eq:Sgeneral} becomes
\begin{align}\label{eq:SKapprox}
	S& =1-(1-\pi)^{s(k-1)}\nonumber\\
	& \quad -\frac{(1-\pi)^{2s(k-1)-2}s(k-1)\pi}{1-(s-1)(k-1)(1-\pi)^{(k-1)s-2}\pi}.
\end{align}
Now $s(k-1)$ is the degree of a vertex. Equation~\eqref{eq:SKapprox} therefore shows that fixing the degree of a vertex, and changing $k$ (by decreasing or increasing $s$) does not influence the leading term for the giant component size $S$. Furthermore, the larger $s$, so the larger the average degree of a vertex, the more dominant the first term becomes. Thus, the larger the degree of a vertex, the smaller the influence of the clique structure of the network on percolation processes.

In particular, fixing the degree of a vertex at $s(k-1)=d$ and investigating the difference between choosing cliques of size $k=i$ or $k=j$ yields
\begin{align}
	S_{K_i}-S_{K_j}& =\frac{d\pi (1-\pi)^{2d-2}}{1-d\pi (1-\pi)^{d-2}+(i-1)\pi(1-\pi)^{d-2}}\nonumber\\
	& \quad -\frac{d\pi (1-\pi)^{2d-2}}{1-d\pi (1-\pi)^{d-2}+(j-1)\pi(1-\pi)^{d-2}}\nonumber\\
	& =O(d\pi^2 (1-\pi)^{3d-4}(j-i)).
\end{align}
Thus, by making $d$ larger, it is always possible to get $S_{K_i}-S_{K_j}$ arbitrarily small. This indeed shows that when the average degree of a network is large, the influence of the clique structure of the model becomes irrelevant. 

\begin{figure*}[tbp]
	\centering
	\subfloat[Degree of each vertex equals 6]{
		\includegraphics[width=0.4\linewidth]{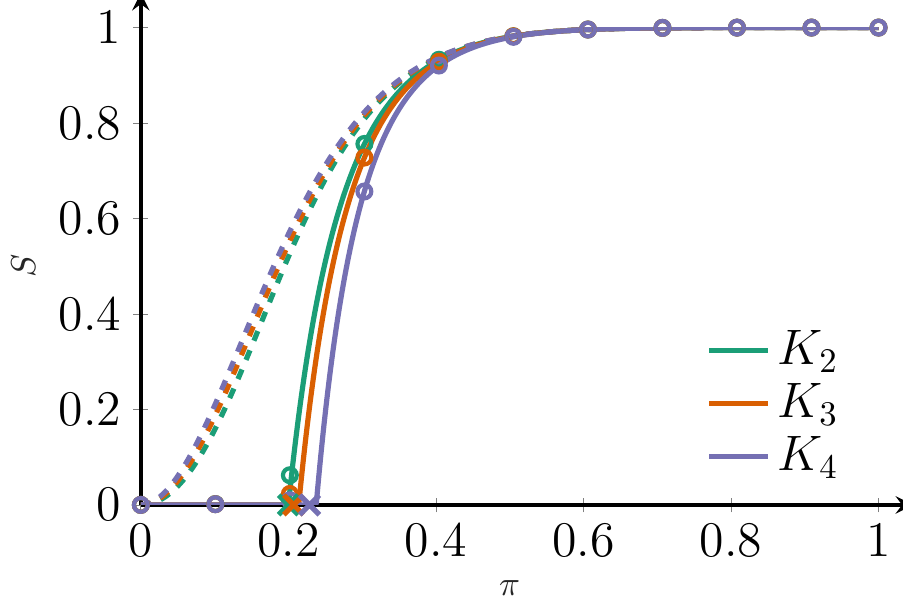}
		\label{fig:regdeglow}
}
\hspace{0.5cm}
	\subfloat[Degree of each vertex equals 12]{
		\includegraphics[width=0.4\linewidth]{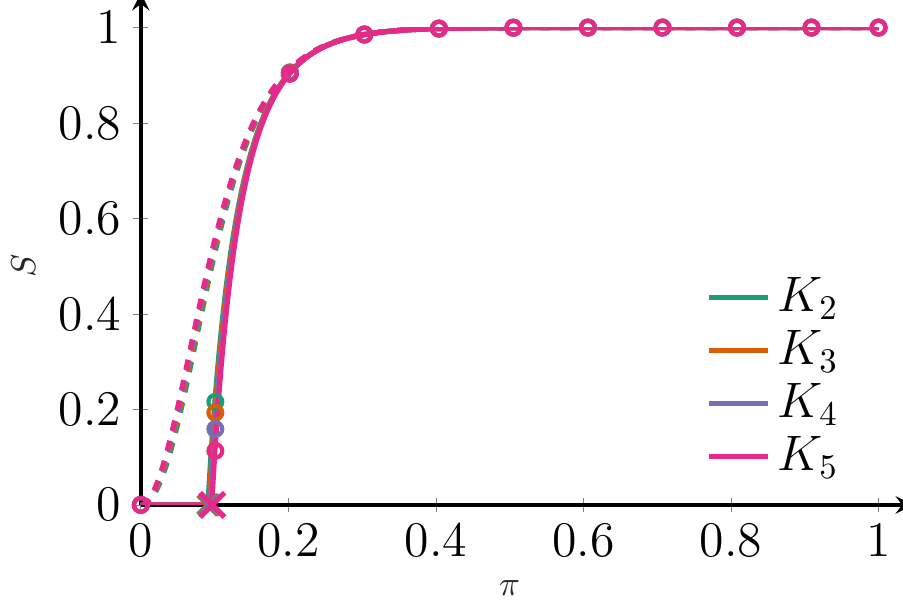}
		\label{fig:regdeghigh}
	}
\caption{Size of the largest component after percolation on networks with only clique-edges of given size. The solid line presents the analytical value of $S$ obtained from solving~\eqref{eq:cliqueuS}, the dashed line is its approximation from~\eqref{eq:SKapprox}, circles are obtained by simulations on $N=10000$ vertices, and the cross indicates the approximation of the critical percolation value from~\eqref{eq:picreg}. }
\label{fig:k4k3}
\end{figure*}

Figure~\ref{fig:k4k3} shows the behavior of the approximation of~\eqref{eq:SKapprox} for three networks, one consisting of only edges (the standard configuration model), one only of triangle-edges, and the other only of $K_4$-edges. We see that the approximation of~\eqref{eq:SKapprox} works well when $S$ is large for all networks. Furthermore, the size of the largest component under percolation differs more between $K_3$ and $K_4$ than between $K_2$ and $K_3$ under small average degree in Figure~\ref{fig:regdeglow}, while these differences have washed away in Fig.~\ref{fig:regdeghigh} under higher average degree.

\paragraph{Example: Poisson degrees}
Under a Poisson degree distribution where every vertex is part on on average $s$ $k$-cliques, the generating functions of~\eqref{eq:cliqueuS} become $g_p(x)=g(x)=e^{s(x-1)}$. 
Then,~\eqref{eq:Sgeneral} becomes
\begin{align}\label{eq:PoissSapprox}
	&S
	 \approx 1-\me^{s((1-\pi)^{k-1}-1)}\nonumber\\
	& \times\Big(1+\frac{s(k-1)\pi(1-\pi)^{2(k-2)}\me^{s((1-\pi)^{k-1}-1)}}{1-s\me^{s((1-\pi)^{k-1}-1)}(k-1)\pi(1-\pi)^{2(k-2)}}\Big).
\end{align}

\begin{figure*}[tbp]
	\centering
\subfloat[Average degree is 6 for all vertices]{
		\centering
		\includegraphics[width=0.4\linewidth]{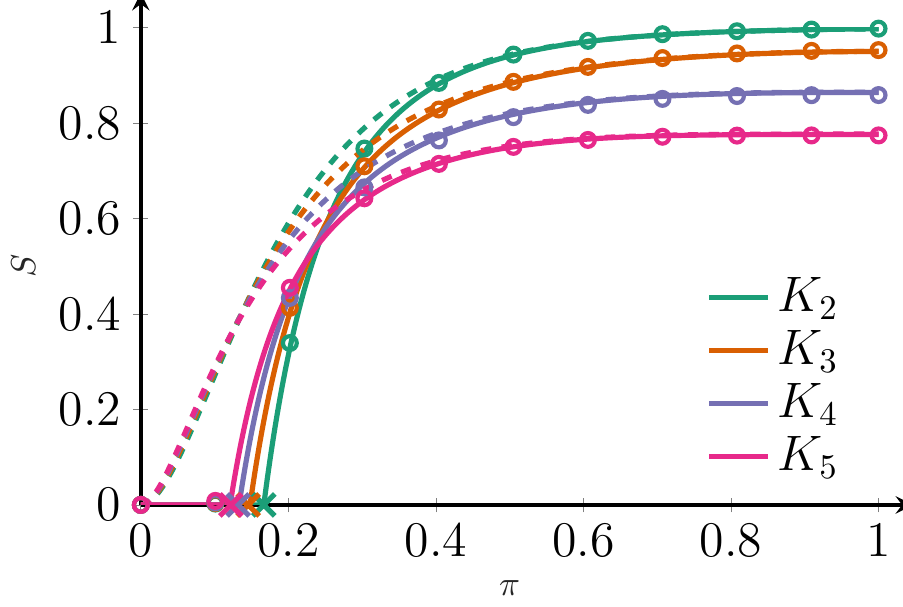}
}
\hspace{0.5cm}
\subfloat[average degree is 12 for all vertices]{
		\centering
		\includegraphics[width=0.4\linewidth]{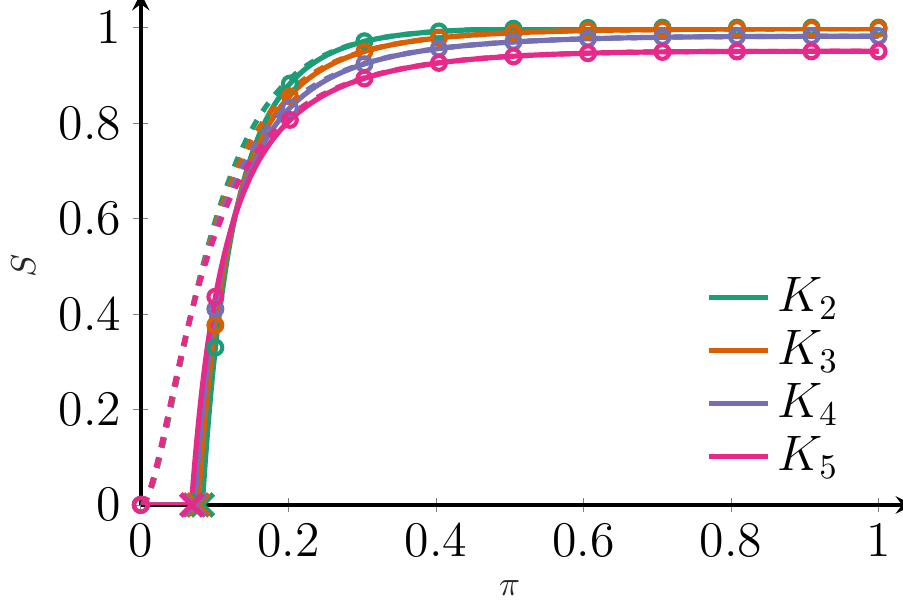}
}
	\caption{Size of the largest component after percolation on Poisson networks with only clique-edges of one specified size. The solid line presents the analytical value of $S$ obtained from solving~\eqref{eq:cliqueuS}, the dashed line is its approximation from~\eqref{eq:PoissSapprox} circles are obtained by simulations on $N=10000$ vertices, and the cross indicates the approximation of the critical percolation value from~\eqref{eq:picpois}.}
	\label{fig:k4k3pois}
\end{figure*}

Figure~\ref{fig:k4k3pois} shows the behavior of the approximation of~\eqref{eq:SKapprox} for four networks, one consisting of only edges, one of only triangle-edges, one of only of $K_4$-edges and one of only $K_5$-edges. We see that for these Poisson degree distributions, the difference between the large outbreak sizes are well approximated by~\eqref{eq:PoissSapprox}, but that these final sizes still differ quite a bit even for large average degrees. This is caused by the fact that for Poisson clique-degrees, the degree distributions of the different clique sizes are \emph{not the same}. 

Indeed, if we focus on the 2-clique case, a vertex can have degree $0,1,2,\dots$ when its degree is sampled from a Poisson degree distribution. However, a vertex that is part of triangles, can only have degrees $0,2,4,6,\dots$, when the number of triangles is sampled from a Poisson distribution. In general, a vertex that is only part of $k$-cliques can only have degrees $0,k-1,2(k-1),\dots$. Even when the average values of the Poisson distributions are tuned as $\lambda/(k-1)$ to make sure that on average, all vertices have the same average number of connections, the degree distributions are not the same. This makes the leading order term in~\eqref{eq:PoissSapprox} different for different clique sizes.  In particular, the probability of having zero connections increases, which makes the final component size smaller when the clique size increases.

\begin{figure}[tbp]
    \centering
    \includegraphics[width=0.45\textwidth]{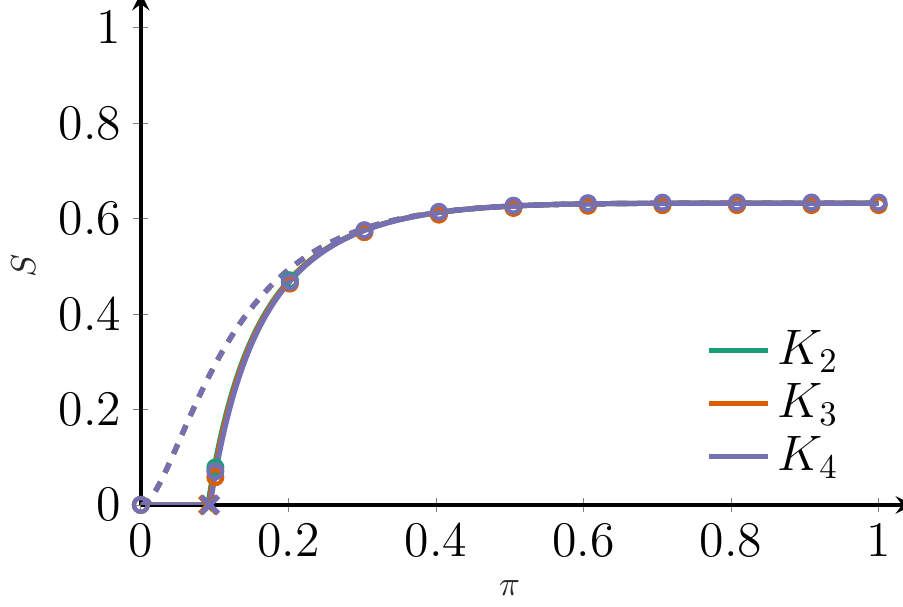}
    \caption{Size of the largest component after percolation on networks with a Poisson degree distribution with $\lambda=1$, adjusted so that every network has the same degree distribution and average degree 6, where every vertex is only part of clique-edges of specified size. The solid line presents the analytical value of $S$ obtained from solving~\eqref{eq:cliqueuS}, the dashed line is its approximation from~\eqref{eq:Sgeneral}, circles are obtained by simulations on $N=10000$ vertices, and the cross indicates the approximation of the critical percolation value from~\eqref{eq:pic}.}
    \label{fig:poisdifnetworks}
\end{figure}

To overcome this problem, we now generate networks with different clique sizes with the same degree distribution. We do this by generating the $K_4$ network by sampling a Poisson random variable for each vertex, which we multiply by 2. This is the $K_4$ degree for each vertex. For the $K_3$ network, we sample a Poisson random variable with the same mean for each vertex, which we multiply by 3. This is the $K_3$ degree for each vertex. For the edge-network we again sample a Poisson random variable with the same mean for each vertex, which we multiply by 6. This is the edge-degree for each vertex. Now, in all three networks, vertices can only have degrees $0,6,12,\dots$, and the degree distribution across the three networks is the same. Figure~\ref{fig:poisdifnetworks} shows the results on percolation on these types of networks. We see that in this case, the percolation curves of these Poisson networks of different clique sizes completely overlap, even while the average degree in this setting is only 6. Thus, the difference between networks of different clique structures under Poisson degree distributions reported in Fig.~\ref{fig:k4k3pois}, but also in~\cite{newman2009,karrer2010}, does in fact not seem to be caused by the clique structure of the network, but by the fact that the degree distributions of the networks are different, changing the leading order term in~\eqref{eq:Sgeneral}.

\begin{figure}[tbp]
    \centering
    \includegraphics[width=0.45\textwidth]{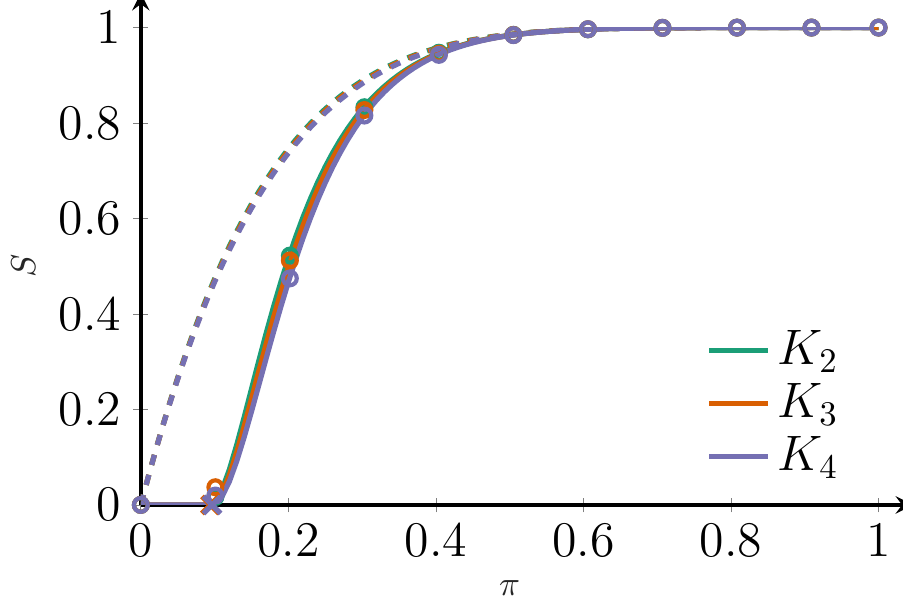}
    \caption{Size of the largest component after percolation on networks with a power-law degree distribution with exponent $\tau=3.5$ and average degree 7.1 where every vertex is only part of clique-edges of a specified size. The solid line presents the analytical value of $S$ obtained from solving~\eqref{eq:cliqueuS}, the dashed line is its approximation from~\eqref{eq:Sgeneral}, circles are obtained by simulations on $N=10000$ vertices, and the cross indicates the approximation of the critical percolation value from~\eqref{eq:pic}.}
    \label{fig:percplnetworks}
\end{figure}

\paragraph{Example: Power-law degrees}
For networks with power-law degrees, we can follow the same approach as for the Poisson networks. We generate power-law random variables, multiply them by 2 for the $K_4$ network, by 3 for the triangle-networks, and by 6 for the edge-network to ensure that all networks have the same degree distribution. Using that $g(x)=\textup{Li}_{\tau}(z)/\zeta(z)$ is the generating function of a power-law random variable with exponent $\tau$, we can again find the approximation of the largest component size under percolation for large outbreaks from~\eqref{eq:Sgeneral}. Figure~\ref{fig:percplnetworks} shows that also for power-law random networks, the large outbreak sizes of the different networks are similar. 

\subsection{Approximation of $\pi_c$ for general clique-degree distributions}\label{sec:piccliques}
We now turn from investigating the similarity of large outbreaks under different clique sizes to investigating the similarity of small outbreaks. In particular, we approximate the critical percolation value $\pi_c$. The critical value $\pi_c$ is obtained when the average number of neighbors  of a vertex reached by following a randomly chosen edge after percolation equals one. Thus,
\begin{equation}
	1=\frac{\langle s^2\rangle-\langle s\rangle}{\langle s\rangle}\sum_{j=1}^{k-1}jh(k,j,\pi_c),
\end{equation}
where $\frac{\langle s^2\rangle-\langle s\rangle}{\langle s\rangle}$ equals the average number of $k$-cliques connected to a vertex reached from an arbitrary $k$-clique

For large average degrees, the critical percolation value is achieved at small $\pi$. Therefore, we only keep terms of order $\pi^2$ or less. The only terms in the summation above with terms of order $\pi^2$ or less are $h(k,1,\pi)$ and $h(k,2,\pi)$, as reaching 3 or more other vertices in a clique requires at least 3 edges to be present, giving a contribution of at least $\pi^3$. By filling in $h(k,1,\pi)=(k-1)\pi(1-\pi)^{2(k-2)}$ and $h(k,2,\pi)=(k-1)(k-2)(3\pi^2(1-\pi)^{3(k-3)+1}+\pi^3(1-\pi)^{3(k-3)})$, we approximate
\begin{align}
	1&\approx \frac{\langle s^2\rangle-\langle s\rangle}{\langle s\rangle}\Big((k-1)\pi_c(1-\pi_c)^{2(k-2)}\nonumber\\
	& \quad +(k-1)(k-2)(3\pi_c^2(1-\pi_c)^{3(k-3)+1}\nonumber\\
	& \quad +\pi_c^3(1-\pi_c)^{3(k-3)})\Big).
\end{align}
Keeping only the terms of order $\pi_c^2$ or less gives
\begin{align}
	1& \approx \frac{\langle s^2\rangle-\langle s\rangle}{\langle s\rangle}(k-1)\left(\pi_c-2(k-2)\pi_c^2+(k-2)3\pi_c^2\right)\nonumber\\
	& = \frac{\langle s^2\rangle-\langle s\rangle}{\langle s\rangle}(k-1)\left(\pi_c+(k-2)\pi_c^2\right).
\end{align}
This is a quadratic equation that has its positive solution at
\begin{equation}\label{eq:pic}
	\pi_c=\frac{-1+\sqrt{1+\frac{4(k-2)}{\frac{\langle s^2\rangle-\langle s\rangle}{\langle s\rangle}(k-1)}}}{2(k-2)}.
\end{equation} 
When the average degree, and therefore also $\frac{\langle s^2\rangle-\langle s\rangle}{\langle s\rangle}(k-1)$, becomes large, we use a first order Taylor expansion of $\sqrt{1+1/x}$ for large $x$. Then, $\pi_c$ can be approximated by
\begin{equation}
    \pi_c\approx \frac{1}{\frac{\langle s^2\rangle-\langle s\rangle}{\langle s\rangle}(k-1)}.
\end{equation}
The term in the denominator describes the average number of vertices reached by coming from a randomly chosen clique-edge, without percolation. When we compare two networks with different clique structures but with the same degree distribution, $\frac{\langle s^2\rangle}{\langle s\rangle}(k-1)$ is the same for the different networks. Furthermore, this quantity is increasing in the average degree. Thus, in the large average degree-regime $\pi_c$ converges to a value that is independent of the clique structure of the network.

\paragraph{Regular networks.} In networks where every vertex is connected to $s$ $k$-cliques, we can reduce~\eqref{eq:pic} in the following way. 
Using that the degree of a vertex $d=s(k-1)$,~\eqref{eq:pic} becomes
\begin{equation}\label{eq:picreg}
	\pi_c=\frac{-1+\sqrt{1+\frac{4(k-2)}{d-(k-1)}}}{2(k-2)}\approx\frac{1}{d-(k-1)}.
\end{equation} 
Thus, when $d$ increases, $\pi_c$ approaches the same value for all cliques sizes $k$. Furthermore, the larger $k$, the larger the difference between $\pi_c$ when increasing $k$ by one. Figure~\ref{fig:k4k3} shows the approximated value of $\pi_c$ from equation~\eqref{eq:picreg} versus the analytical values of the giant component sizes. We see that already for an average degree of 6 this approximation is quite good, and that for larger average degree of 12, indeed the values of $\pi_c$ for the network of triangles and $K_4$ cliques almost overlap.

\paragraph{Poisson networks.}
In Poisson networks where the average vertex is part of $s$ $k$-cliques with fixed $d=s(k-1)$, $\frac{\langle s^2\rangle-\langle s\rangle}{\langle s\rangle}=s$. Then~\eqref{eq:pic} becomes 
\begin{equation}\label{eq:picpois}
	\pi_c=\frac{-1+\sqrt{1+\frac{4(k-2)}{d}}}{2(k-2)}\approx\frac{1}{d}.
\end{equation} 
Thus, when $d$ gets large, again $\pi_c$ approaches the same value for all cliques sizes $k$. Figure~\ref{fig:k4k3pois} shows that this is a good approximation of the critical percolation value $\pi_c$, and that for an average degree of 12, these values become very close under different clique sizes.


\begin{figure}[t!]
    \centering
    \includegraphics[width=0.45\textwidth]{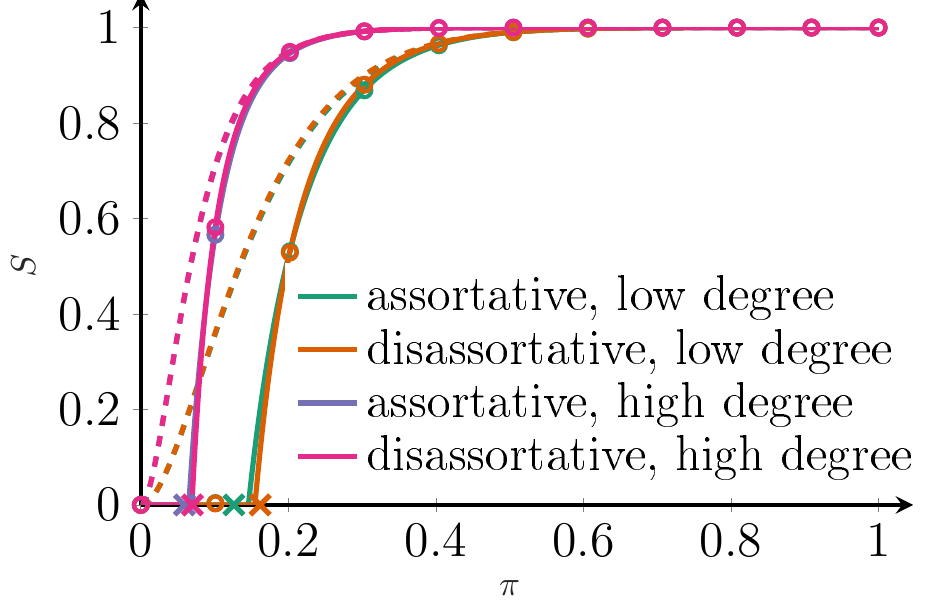}
    \caption{Size of the largest component after percolation on networks with a clique sizes of 2 and 4, mixed assortatively or disassortatively with average degree 7.5 or 15. (here $p_{i,j}$ denotes the probability of having $i$ cliques of size 2 and $j$ of size 4, and for assortative $p_{6,0}=0.5,p_{3,2}=0.25, p_{0,3}=0.25$, while for disassortative $p_{3,1}=0.5,p_{3,2}=0.5$. In the high-degree regime, all degrees are doubled. Dashed lines are the approximations from~\eqref{eq:Smixed}, and the cross denotes the approximation of $\pi_c$ from~\eqref{eq:picmixed1}.}
    \label{fig:percmixed}
\end{figure}

\section{Mixed clique sizes}\label{sec:mixedcliques}
We now investigate networks where cliques of different sizes are present. By introducing different clique sizes, it is possible to create degree-degree correlations that have often been said to influence the largest component size after percolation~\cite{boguna2002,goltsev2008,balogh2020}. Thus, we now investigate to what extent the introduction of mixed clique sizes influences the size of the largest component.

Under bond percolation of networks where every vertex is part of $s_1$ cliques of size $k_1$, and $s_2$ cliques of size $s_2$ with probability $q_{s_1,s_2}$ the generating-function methodology gives the following results. 
Let $g(x,y)=\sum_{s_1,s_2>0}p_{s_1,s_2}x^{s_1}y^{s_2}$ be the generating function of the clique degrees.
Furthermore, let
\begin{align}
	g_p(x,y)& =\frac{1}{\langle s_1\rangle}\sum_{s_1,s_2>0}{s_1}q_{s_1,s_2}x^{s_1-1}y^{s_2},\\
	g_q(x,y)& =\frac{1}{\langle s_2\rangle}\sum_{s_1,s_2>0}{s_2}q_{s_1,s_2}x^{s_1}y^{s_2-1},
\end{align}
with
$$
\langle s_1\rangle := \sum_{s_1,s_2>0}s_1q_{s_1,s_2},\quad\langle s_2\rangle:=\sum_{s_1,s_2>0}lq_{s_1,s_2},
$$
be the generating functions of the number of cliques that are reached by following a randomly chosen clique-edge.
Let $u$ denote the probability that a randomly chosen $k_1$-clique-edge is not connected to the giant component. Similarly, let $v$ denote the probability that following a randomly chosen $k_2$-clique edge does not lead to the largest component.

We show in Appendix~\ref{sec:mixedcliquecomps} that $u$ and $v$ can be approximated by
\begin{align}
	u&\approx\frac{g_p((1-\pi)^{k_1-1},(1-\pi)^{k_2-1})}
	{A(k_1,k_2,\pi)}\nonumber\\
	v&\approx \frac{g_q((1-\pi)^{k_1-1},(1-\pi)^{k_2-1})}{A(k_1,k_2,\pi)}.
\end{align}
Furthermore, the giant component size $S$ is then approximated as
\begin{align}\label{eq:Smixed}
	S =&1-g_D(1-\pi)-\pi(1-\pi)^{2(k_1-2)}(k_1-1)\nonumber\\
	&\times \frac{g_p((1-\pi)^{k_1-1},(1-\pi)^{k_2-1})^2\langle s_1\rangle }
	{A(k_1,k_2,\pi)}\nonumber\\
	& 	-\pi(1-\pi)^{2(k_2-2)}(k_2-1)\nonumber\\
	& \times \frac{g_q((1-\pi)^{k_1-1},(1-\pi)^{k_2-1})^2\langle s_2\rangle}
	{A(k_1,k_2,\pi)},
\end{align}
where
\begin{align*}
	&A(k_1,k_2,\pi) =\nonumber \\ & 1-(k_1-1)\pi(1-\pi)^{2(k_1-1)}\frac{\partial g_p((1-\pi)^{k_1-1},(1-\pi)^{k_2-1})}{\partial x}\nonumber\\
	& -(k_2-1)\pi(1-\pi)^{2(k_2-1)}\frac{\partial g_q((1-\pi)^{k_1-1},(1-\pi)^{k_2-1})}{\partial y},
\end{align*}
and where $g_D(x)$ is the generating function of the total vertex degrees. Thus, the leading order term of the giant component size does not depend on the distribution of the clique degrees $k_1$ and $k_2$, but only on the total vertex degree. Furthermore, the numerators of the second order terms also only depend on the degree distribution, and not on the clique sizes, similarly to the one clique-size case.

It is not difficult to extend this analysis to include more than two different clique sizes, where~\eqref{eq:Smixed} contains terms for all size biased generating functions of the clique sizes $g_{k_i}$, instead of only $g_p$ and $g_q$ in~\eqref{eq:Smixed}. Therefore, even in the presence of multiple clique sizes that can generate degree-degree correlations, large outbreaks are clique-structure independent for large average degrees.

\paragraph*{Example: Assortative mixing.}
In several sources of previous work, degree-degree correlations were found to be important for the behavior of percolation processes. Furthermore, the clustering assortativity, describing the tendency of high-degree vertices to be more clustered than high-degree vertices or vice versa, has also been ascribed strong importance on the behavior of a network under percolation~\cite{mann2021random}. However,~\eqref{eq:Smixed} shows that large outbreaks only depend on the degree distribution, so that it is independent of any clique correlations in the large degree limit. Figure~\ref{fig:percmixed} shows that indeed the influence of mixed clique sizes on the giant outbreak is small, especially in the large average degree regime.

\subsection{Approximation of $\pi_c$ for mixed clique networks}
In Appendix~\ref{sec:mixedcliquecomps} we show that $\pi_c$ can be approximated by 
\begin{widetext}
\begin{equation}\label{eq:picmixed1}
	\pi_c=\frac{-E_{k_1,k_1}-E_{k_2,k_2}+\sqrt{(E_{k_1,k_1}+E_{k_2,k_2})^2-4\left(E_{k_1,k_1} E_{k_2,k_2}- E_{k_1,k_2} E_{k_2,k_1}-E_{k_1,k_1}(k_1-2)-E_{k_2,k_2}(k_2-2)\right)}}{2\left(E_{k_1,k_1} E_{k_2,k_2}- E_{k_1,k_2} E_{k_2,k_1}-E_{k_1,k_1}(k_1-2)-E_{k_2,k_2}(k_2-2)\right)},
\end{equation}
\end{widetext}
where 
\begin{equation}
    E_{k_i,k_j}=\left(\frac{\langle s_is_j\rangle}{\langle s_j\rangle}-\delta_{k_i,k_j}\right)(k_i-1).
\end{equation}
For large average degrees, this value can be approximated by 
\begin{align}
	\pi_c & \approx \frac{1}{E_{k_1,k_1}+E_{k_2,k_2}}\nonumber\\
	& =\frac{1}{\left(\frac{\langle s_1^2\rangle}{\langle s_1\rangle}-1\right)(k_1-1)+\left(\frac{\langle s_2^2\rangle}{\langle s_2\rangle}-1\right)(k_2-1)}.
\end{align}
In assortative networks, where cliques of a given size are typically also connected to many cliques of the same size, $\left(\frac{\langle s_1^2\rangle}{\langle s_1\rangle}-1\right)(k_1-1)$ and $\left(\frac{\langle s_2^2\rangle}{\langle s_2\rangle}-1\right)(k_2-1)$ are large, so that we expect $\pi_c$ to be small. In disassortative networks, where the different clique sizes are more mixed, $\left(\frac{\langle s_1^2\rangle}{\langle s_1\rangle}-1\right)(k_1-1)$ and $\left(\frac{\langle s_2^2\rangle}{\langle s_2\rangle}-1\right)(k_2-1)$ are smaller. Thus, the degree-degree correlations that are created by the different clique sizes play a role in the critical percolation value $\pi_c$, whereas the giant outbreak size of~\eqref{eq:Smixed} is asymptotically independent of such degree correlations. However, Figure~\ref{fig:percmixed} shows that these correlations still vanish in the large-degree regime.

\section{Phase oscillators coupled on clique-networks}\label{sec:Kuramoto}

In this section we illustrate the limited effect of higher-order structure on more complex dynamic network processes than bond percolation. In particular, we focus on the dynamics of coupled oscillators. For this purpose, we employ the paradigmatic Kuramoto model~\cite{rodrigues2016kuramoto} that can describe synchronization phenomena on complex networks. In the Kuramoto model, the oscillator 
of vertex $i$ is characterized by a phase variable $\theta_i$, and the dynamics on a heterogeneous network is dictated by the following equations~\cite{rodrigues2016kuramoto}: 
\begin{equation}
\frac{d\theta_i}{dt} = \omega_i + K \sum_{j=1}^N A_{ij} \sin(\theta_j - \theta_i), \; i=1,...,N, 
\label{eq:KM_model}
\end{equation}
where $\omega_i$ is the natural frequency of oscillation of oscillator $i$, and $A_{ij}$ is the network adjacency matrix. If there is an edge connecting $i$ and $j$, $A_{ij} = 1$ (0 otherwise), and the interaction between the vertices is weighted by the coupling $K$. If $K$ is lower than a certain $K_c$, the oscillators rotate incoherently, each one at its own rhythm set by the natural frequency $\omega_i$. For $K>K_c$, the incoherent state loses stability: a cluster of oscillators is formed around an average phase value, and these units begin to rotate locked in the same frequency
~\cite{strogatz2000kuramoto,fonseca2018kuramoto}. This transition from asynchrony to a partially synchronized state is measured by the Kuramoto order parameter $R$ given by~\cite{strogatz2000kuramoto,fonseca2018kuramoto} 
\begin{equation}
R e^{i\psi}= \frac{1}{N} \sum_{j=1}^N e^{i\theta_j} \quad (0\leq R \leq 1),
\label{eq:order_parameter} 
\end{equation} 
where $R$ quantifies the level of synchrony achieved by the oscillators, and $\psi$ is their average phase. While one can monitor the synchronization transition of a heterogeneous network with Eq.~(\ref{eq:order_parameter}), it is not possible to decouple Eqs.~\eqref{eq:KM_model} in terms of a global order parameter. Instead, in order to perform a self-consistent analysis and characterize the onset of synchronization analytically, we need to employ heterogeneous degree mean-field approximations
~\cite{rodrigues2016kuramoto}. This is equivalent to replacing the terms of the adjacency matrix $A_{ij}$ by their ensemble averages in the configuration model, which in the single-edge version is $\langle A_{ij} \rangle = d_i d_j/N\langle d \rangle$. In the model that generates networks with a single clique type the expression is analogous, namely, $\langle A_{ij}^{(c)} \rangle = (k-1) c_i c_j /N\langle c \rangle$, where $c_i$ is the number of cliques attached to vertex $i$. Replacing $A_{ij}$ by $\langle A_{ij}^{(c)} \rangle$ in Eq.~(\ref{eq:KM_model}) we obtain
\begin{equation}
    \frac{d\theta_i}{dt} = \omega_i + \frac{K(k-1)c_i}{N\langle q \rangle} \sum_{j=1}^N c_j \sin(\theta_j - \theta_i), \; i=1,...,N, 
    \label{eq:KM_model_c}
\end{equation}
which motivates the definition of the following order parameter
\begin{equation}
    r e^{i\phi} = \frac{1}{N \langle c \rangle} \sum_{j=1}^N c_j e^{i\theta_j},
    \label{eq:order_parameter_hmf}
\end{equation}
which in turn allows us to rewrite Eq.~(\ref{eq:KM_model_c}) as
\begin{equation}
    \frac{d\theta_i}{dt} = \omega_i + Kr(k-1)c_i\sin(\phi - \theta_i).
    \label{eq:KM_net_decoupled}
\end{equation}
In the limit of $N \rightarrow \infty$, we assume that the assignment of cliques and natural frequencies is well described by distributions $q_c$ and $g(\omega)$; we further assume that the collections of vertices with clique number $c$ and frequency $\omega$ form a phase density $\rho(\theta,t|c,\omega)$. Thus, we rewrite Eq.~\eqref{eq:order_parameter_hmf} in the continuum limit as
\begin{equation}
    r e^{i\phi} = \frac{1}{\langle c \rangle} \sum_c c q_c \int \int d\omega d\theta \rho(\theta,t|c,\omega) g(\omega) e^{i\theta}
    \label{eq:order_parameter_hmf_cont}
\end{equation}
By choosing $g(\omega) = (\sqrt{2\pi})^{-1} e^{-\omega^2/2}$, we can set $\phi=0$ without loss of generality. Substituting the stationary synchronous solution of Eq.~(\ref{eq:KM_net_decoupled})--i.e. $\rho(\theta|\omega) = \delta\{\theta - \arcsin[\omega/Kr(k-1)c]\}$, for $|\omega| \leq Kr(k-1)c$--into Eq.~(\ref{eq:order_parameter_hmf_cont}), we arrive at the following implicit equation
\begin{align}
\frac{\langle c\rangle}{K}\sqrt{\frac{8}{\pi}} & =(k-1)\sum_{c}c^{2}q_{c}e^{-K^{2}(k-1)^{2}c^{2}r^{2}/4}\nonumber\\
\times & \left\{ I_{0}\left[\frac{K^{2}(k-1)^{2}c^{2}r^{2}}{4}\right]+I_{1}\left[\frac{K^{2}(k-1)^{2}c^{2}r^{2}}{4}\right]\right\}, \label{eq:implicit_eq}
\end{align}

where $I_0(\cdot)$ and $I_1(\cdot)$ are the modified Bessel functions of the first kind. Thus, with Eq.~(\ref{eq:implicit_eq}) we can find the dependence of the order parameter $r$ on the coupling strength $K$, and thereby assess the impact of different clique sizes on the onset of synchronization. Letting $r\rightarrow 0^+$, we also obtain the expression for the critical coupling
\begin{equation}
    K_c = \frac{1}{(k-1)}  \frac{\langle c \rangle}{\langle c^2 \rangle} \sqrt{\frac{8}{\pi}}.
    \label{eq:critical_coupling}
\end{equation}
Thus, again, plugging in $d=(k-1)c$ shows that the critical coupling is independent of the clustering structure. 
However, an immediate problem we face is the fact that the mean field approximations behind Eq.~(\ref{eq:implicit_eq}) are accurate only for sufficiently dense networks, typically when the average degree is at least of order of a few dozen~\cite{rodrigues2016kuramoto,gleeson2012accuracy}. This limits the analytical verification of the effect of cliques on the dynamics of networks as sparse as the ones considered in the previous sections. Nonetheless, in the appropriate regime in which the mean field approach is valid, Eq.~(\ref{eq:implicit_eq}) suggests that the conclusions drawn for bond percolation may be similar for synchronization processes: Notice that in Eq.~(\ref{eq:implicit_eq}) $(k-1)c$ is the actual degree of a vertex; substituting $d = (k-1)c$ in the implicit equation for $r$ and rewriting it in terms of the new variable, we find that the emergence of a synchronous component depends only on the final degree sequence of the network and not on the sizes of the cliques. Therefore, clustered and unclustered networks are expected to exhibit similar dynamics also in the synchronization of coupled oscillators. 

\begin{figure}[t!]
    \centering
    \includegraphics[width=0.45\textwidth]{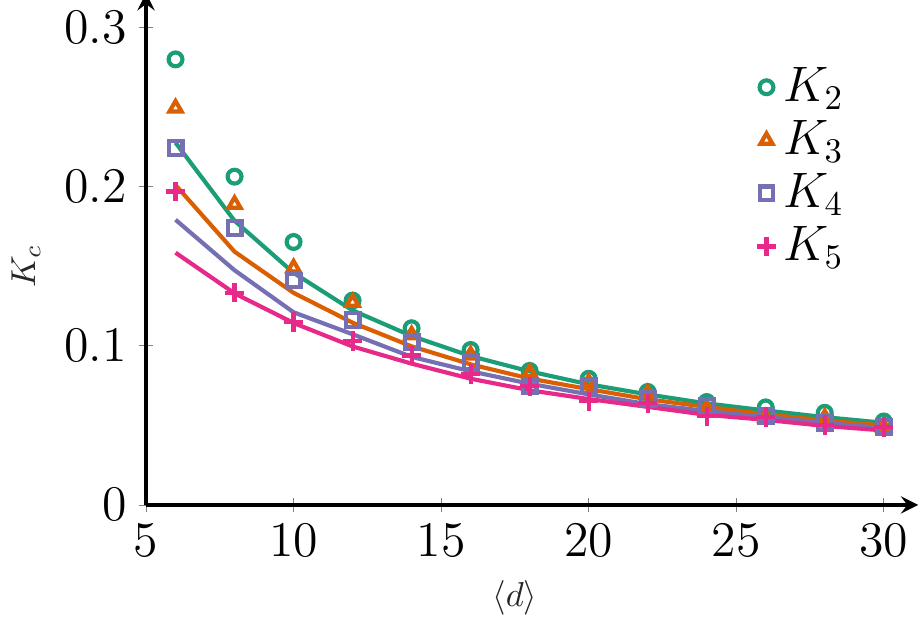}
    \caption{Critical coupling $K_c$ for the onset of synchronization as a function of the average degree $\langle d \rangle$ for the clique-networks with Poisson degree distributions. Dashed lines are the solutions of Eq.~(\ref{eq:critical_coupling}). Dots correspond to the simulation results obtained by numerically integrating Eq.~(\ref{eq:KM_model}) using the Heun's method with time step $dt=0.05$. For each coupling $K$, the quantities are average over $t \in [500,1000]$. In all numerical experiments we have $N=10^4$ oscillators, whose frequencies are distributed according to $g(\omega) = (\sqrt{2\pi})^{-1} e^{-\omega^2/2}$. Different symbols and colors refer to networks constructed with different clique sizes: $K_2$ denotes networks containing only single-edges (configuration model), and $K_5$ refers to networks built from sequences of cliques with five vertices.}
    \label{fig:sync_Kc}
\end{figure}

In order to confirm the above result, let us first investigate how the critical point $K_c$ changes according to the clique structure. In Fig.~\ref{fig:sync_Kc} we compare the predictions of $K_c$ by Eq.~(\ref{eq:critical_coupling}) with the corresponding quantities obtained via numerical integration of the system (\ref{eq:KM_model}) for several average degrees $\langle d \rangle$. We numerically detect the transition point between incoherence and partial synchronization by identifying $K_c$ as the position of the divergent peak of the susceptibility $\chi = N(\langle r^2 \rangle_t - \langle r \rangle_t )/\langle r \rangle_t$~\cite{peron2019onset}, where $\langle \cdot \rangle_t$ is a long temporal average. As can been in Fig.~\ref{fig:sync_Kc}, the agreement between simulation and theoretical values is satisfactorily good for low $\langle d \rangle$, but it is progressively improved as the networks get denser. Furthermore, Fig.~\ref{fig:sync_Kc} indicates that the transition to synchrony tends to occur sooner as the clique size, and hence the clustering, increases. The numerical value of $K_c$ for different clique sizes becomes statistically equivalent at high $\langle d \rangle$. Yet, the solutions obtained from Eq.~\eqref{eq:critical_coupling} in Fig.~\ref{fig:sync_Kc} suggest that clustering always ameliorates the network synchrony, as seen in Fig.~\ref{fig:sync_Kc}, an effect that asymptotically vanishes as $\langle d \rangle$ increases. This is in apparent contradiction with our analysis of Eq.~(\ref{eq:implicit_eq}), in that networks with the same degree distribution, regardless of their clustering structure, ought to have identical dependence $r=r(K)$ and critical couplings $K_c$. However, similarly to the experiments of Fig.~\ref{fig:k4k3pois}, these networks of different clique sizes do not have the same degree distributions.

\begin{figure}[t!]
    \centering
    \includegraphics[width=0.45\textwidth]{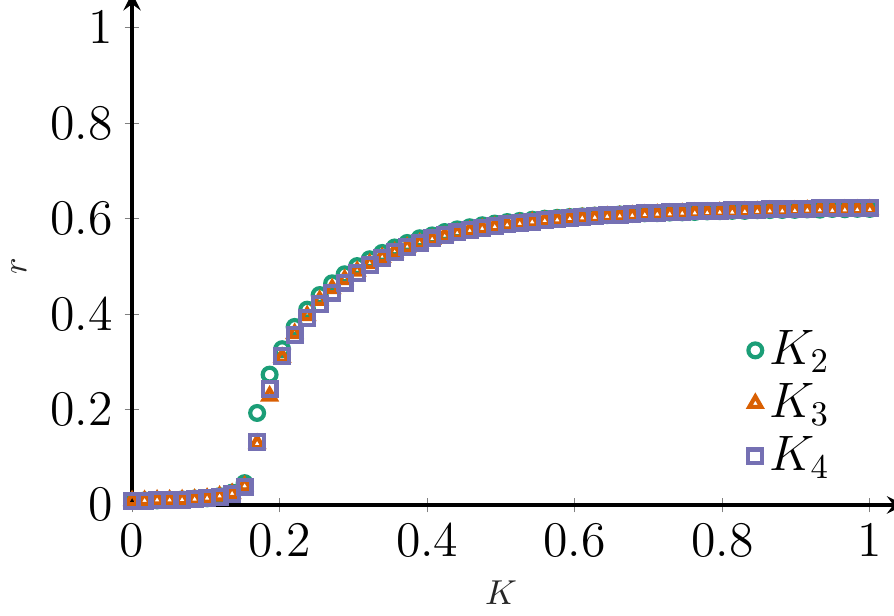}
    \caption{Synchronization diagram showing the evolution of the order parameter $r$ as a function of the coupling $K$. The networks have Poisson degree distribution adjusted so that every network has the same degree sequence and average degree $\langle d \rangle = 6$, as in Fig.~\ref{fig:poisdifnetworks}. Dots correspond to the simulation results obtained by numerically integrating Eq.~(\ref{eq:KM_model}) using the Heun's method with time step $dt=0.05$. For each coupling $K$, the order parameters are average over $t \in [500,1000]$. In all numerical experiments we have $N=10^4$ oscillators, and the frequencies are distributed according to $g(\omega) = (\sqrt{2\pi})^{-1} e^{-\omega^2/2}$. Different symbols and colors refer to networks constructed with different clique sizes: $K_2$ denotes networks containing only single-edges (configuration model), and $K_5$ refers to networks built from sequences of cliques with five vertices.}
    \label{fig:sync_ER_special}
\end{figure}

To verify whether the differences in the onset of synchronization shown in Fig.~\ref{fig:sync_Kc} are due to discrepancies in the degree sequences or are, on the other hand, a true effect of the clustered structure, we repeat the methodology of the experiments depicted in Fig.~\ref{fig:poisdifnetworks}. That is, we simulate the oscillators on networks with different clique sizes but adjusted to have the exact same degree sequence. The result is seen Fig.~\ref{fig:sync_ER_special}. Again, as in the percolation experiment in Fig.~\ref{fig:poisdifnetworks}, the synchronization values match almost perfectly despite the difference in the clustering levels: in the examples in Fig.~\ref{fig:sync_ER_special}, the single-edge networks ($K_2$) have transitivity coefficient~\cite{newman2009} equal to $C \approx 0$, while networks with $K_4$ cliques exhibit $C \approx 0.18$--a significant structural difference that is not reflected in the dynamics. It is also noteworthy that this phenomenon occurs at a low average degree ($\langle d \rangle = 6$), i.e., the regime depicted in Fig.~\ref{fig:sync_Kc} with the most prominent discrepancies between 
clustered and unclustered networks. As discussed for the percolation case in Sec.~\ref{sec:perccliques}, actually those discrepancies are due the fact that the degree distributions are not identical for different clique sizes; as a consequence, the mismatches in the degree sequence end up generating different solutions for Eq.~(\ref{eq:implicit_eq}). The results in Figs.~\ref{fig:poisdifnetworks} and ~\ref{fig:sync_ER_special} therefore show that networks with similar degree distributions and correlations may exhibit equivalent dynamical behavior regardless of their subgraph structure and clustering levels.   

We also note that the critical coupling $K_c$ can be estimated via ``quenched'' mean-field approximations and, for the parameters considered here, expressed in terms of the largest eigenvalue $\lambda_1$ of the adjacency matrix as $K_c = \lambda_1^{-1}\sqrt{8/\pi} $~\cite{restrepo2005onset,peron2019onset}. We can complement the latter expression with the recent
results of Ref.~\cite{peron2018spectra} in which 
the largest eigenvalue for Poisson random networks constructed with $K_3$ cliques has been estimated as $\lambda_1 = 2\langle c \rangle + 1 + 1/\langle c \rangle$. In the limit of high average degrees, $\langle c \rangle \rightarrow \infty$, the third term of $\lambda_1$ vanishes, and the corresponding result for tree-like Poisson random networks is recovered ($\lambda_1 = \langle d \rangle + 1$). Therefore, also in the quenched mean-field formulation, the value of $K_c$ of clustered networks is expected to asymptotically approach the calculations for tree-like networks, in agreement with the results in Fig.~\ref{fig:sync_Kc}.

\section{Conclusion}

In this paper, we have investigated the influence of the presence of clustered structures in random graphs in the form of cliques on two network processes: bond percolation and synchronization. Percolation on such clustered networks has been investigated frequently, but as the equations for the giant component size under percolation are given by several implicit equations that are difficult to analyze mathematically, the factors dominating the behavior of percolation processes on such networks are largely unknown. By approximating the size of the giant component under large outbreaks as well as the critical percolation value where a giant component starts to form, we have found that the degree distribution is the dominant factor in these approximations, especially when the average degree of the network is large. In particular, our approximations are independent of the amount of clustering in the network. This means that introducing clustering by locally inserting cliques or other types of subgraphs in the frequently used locally tree-like random graph models, barely influences the size of the largest component.

We also show that differences in percolation behavior due to the introduction of cliques in the configuration model that were found in previous works can be ascribed to the fact that the degree distribution changed in those experiments as well. When keeping the degree distribution fixed while introducing more clustering, this difference disappears.

While our approximations show that the dominant factor for large outbreaks as well as the critical percolation value is the degree distribution, and not the clustering in the network, our simulations show that actually the entire percolation curve seems to become independent of the clustering in the network once the average degree becomes large. Showing this analytically would be an interesting point for further research.

Furthermore, while we have primarily focused on the process of bond percolation, we also showed that for a different network process of oscillator synchronization, the same independence of higher-order structures is present when the average degree is large. We therefore believe that other processes such as opinion dynamics or the contact process could be independent of the clustering structure of this model as well. Investigating which types of dynamics are independent of the clique structures is therefore an interesting avenue for further research.

Another interesting line of research following from these results is in higher-order dynamics. We showed that singe-edge dynamics on networks where a clique structure is imposed behave similarly as in networks without the clique structure. However, when studying the network model for example as a simplicial complex instead, it is possible to impose simplicial dynamics on top of it, where the dynamics involve all clique vertices in the interactions. It would be interesting to see under which conditions on the dynamic process on such a simplical extension of this model depends on the clique structure, and under which conditions it does not.

Finally, while this work shows that inserting clustering in a locally tree-like model barely affects the behavior of an epidemic process under bond percolation, we believe that in different models, where clustering is introduced by the presence of geometry, bond percolation can behave very differently under two models of the same degree distribution. Showing general conditions on the network structure under which clustering does or does not affect the size of a giant component compared to tree-like network models would therefore also be an interesting avenue for further research. 

\section{Acknowledgments}
T.P. acknowledges FAPESP (grant No. 2016/23827-6).

\bibliographystyle{abbrv}
\bibliography{refs}

 \appendix
 
 \section{Computations for the mixed clique sizes}\label{sec:mixedcliquecomps}
After bond percolation with probability $\pi$~\cite{newman2009}, 
\begin{align}\label{eq:percmixed}
	u& =g_p(\sum_{j=1}^{k_1-1}h(k_1,j,\pi)u^j,\sum_{j=1}^{k_2-1}h(k_2,j,\pi)v^j),\nonumber\\
	v& =g_q(\sum_{j=1}^{k_1-1}h(k_1,j,\pi)u^j,\sum_{j=1}^{k_2-1}h(k_2,j,\pi)v^j),
\end{align}
while $S=1-g(\sum_{j=1}^{k_1-1}h(k_1,j,\pi)u^j,\sum_{j=1}^{k_2-1}h(k_2,j,\pi)v^j)$.
Again, we expand~\eqref{eq:percmixed} with a first-order Taylor expansion around $u,v=0$. This yields
\begin{align}
	u&=g_p\Big((1-\pi)^{(k_1-1)}+(k_1-1)\pi(1-\pi)^{2(k_1-2)}u,\nonumber\\
	& (1-\pi)^{(k_2-1)}+(k_2-1)\pi(1-\pi)^{2(k_2-2)}v\Big)\nonumber\\
	v&=g_q\Big((1-\pi)^{(k_1-1)}+(k_1-1)\pi(1-\pi)^{2(k_1-2)}u,\nonumber\\
	& (1-\pi)^{(k_2-1)}+(k_2-1)\pi(1-\pi)^{2(k_2-2)}v\Big)
\end{align}
Taylor expanding $g_p(x,y)$ and $g_q(x,y)$ as well and using that $uv$ is small, we obtain
\begin{align}
	&u=g_p((1-\pi)^{k_1-1},(1-\pi)^{k_2-1})\nonumber\\
	& +\frac{\partial g_p((1-\pi)^{k_1-1},(1-\pi)^{k_2-1})}{\partial x}(k_1-1)\pi (1-\pi)^{2(k_1-2)}u\nonumber\\
	&  + \frac{\partial g_p((1-\pi)^{k_1-1},(1-\pi)^{k_2-1})}{\partial y}(k_2-1)\pi (1-\pi)^{2(k_2-2)}v\nonumber\\
	&v=g_q((1-\pi)^{k_1-1},(1-\pi)^{k_2-1})\nonumber\\
	&  +\frac{\partial g_q((1-\pi)^{k_1-1},(1-\pi)^{k_2-1})}{\partial x}(k_1-1)\pi (1-\pi)^{2(k_1-2)}u\nonumber\\
	&  + \frac{\partial g_q((1-\pi)^{k_1-1},(1-\pi)^{k_2-1})}{\partial y}(k_2-1)\pi (1-\pi)^{2(k_2-2)}v\nonumber\\
\end{align}
This is a linear system of equations with as solution
\begin{widetext}
\begin{align}
	u&=\frac{g_p((1-\pi)^{k_1-1},(1-\pi)^{k_2-1})}{A(k_1,k_2,\pi)}-\frac{g_p((1-\pi)^{k_1-1},(1-\pi)^{k_2-1})(k_2-1)\pi(1-\pi)^{2(k_2-1)}\frac{\partial g_q((1-\pi)^{k_1-1},(1-\pi)^{k_2-1})}{\partial y}}{A(k_1,k_2,\pi)}\nonumber\\
	& \quad +\frac{g_q((1-\pi)^{k_1-1},(1-\pi)^{k_2-1})(k_2-1)\pi(1-\pi)^{2(k_2-1)}\frac{\partial g_q((1-\pi)^{k_1-1},(1-\pi)^{k_2-1})}{\partial y}}
	{A(k_1,k_2,\pi)}\nonumber\\
		v&=\frac{g_q((1-\pi)^{k_1-1},(1-\pi)^{k_2-1})}{A(k_1,k_2,\pi)} +\frac{g_p((1-\pi)^{k_1-1},(1-\pi)^{k_2-1})(k_1-1)\pi(1-\pi)^{2(k_1-1)}\frac{\partial g_p((1-\pi)^{k_1-1},(1-\pi)^{k_2-1})}{\partial x}}{A(k_1,k_2,\pi)}\nonumber\\
		& \quad 
			-\frac{g_q((1-\pi)^{k_1-1},(1-\pi)^{k_2-1})(k_1-1)\pi(1-\pi)^{2(k_1-1)}\frac{\partial g_q((1-\pi)^{k_1-1},(1-\pi)^{k_2-1})}{\partial x}}
		{A(k_1,k_2,\pi)},
\end{align}
\end{widetext}
where
\begin{align}
	& A(k_1,k_2,\pi)  =\nonumber\\
	& 1-(k_1-1)\pi(1-\pi)^{2(k_1-1)}\frac{\partial g_p((1-\pi)^{k_1-1},(1-\pi)^{k_2-1})}{\partial x}\nonumber\\
	& -(k_2-1)\pi(1-\pi)^{2(k_2-1)}\frac{\partial g_q((1-\pi)^{k_1-1},(1-\pi)^{k_2-1})}{\partial y}.
\end{align}
This can be approximated by
\begin{align}
	u&\approx\frac{g_p((1-\pi)^{k_1-1},(1-\pi)^{k_2-1})}
	{A(k_1,k_2,\pi)}\nonumber\\
	v&\approx \frac{g_q((1-\pi)^{k_1-1},(1-\pi)^{k_2-1})}{A(k_1,k_2,\pi)}.
\end{align}

This gives for the final component size
\begin{align}
	S&=1-g((1-\pi)^{k_1-1},(1-\pi)^{k_2-1})\nonumber\\
	& \quad -\pi(1-\pi)^{2(k_1-2)}(k_1-1)\nonumber\\
	& \times\frac{g_p((1-\pi)^{k_1-1},(1-\pi)^{k_2-1})\frac{\partial g((1-\pi)^{k_1-1},(1-\pi)^{k_2-1})}{\partial x}}
	{A(k_1,k_2,\pi)}\nonumber\\
	& \quad 	-\pi(1-\pi)^{2(k_2-2)}(k_2-1)\nonumber\\
	& \times \frac{g_q((1-\pi)^{k_1-1},(1-\pi)^{k_2-1})\frac{\partial g((1-\pi)^{k_1-1},(1-\pi)^{k_2-1})}{\partial y}}
	{A(k_1,k_2,\pi)}
\end{align}

This can be written as
\begin{align}
	S& =1-g_D(1-\pi)-\pi(1-\pi)^{2(k_1-2)}(k_1-1)\nonumber\\
	& \times \frac{g_p((1-\pi)^{k_1-1},(1-\pi)^{k_2-1})^2\langle s_1\rangle }
	{A(k_1,k_2,\pi)}\nonumber\\
	& \quad 	-\pi(1-\pi)^{2(k_2-2)}(k_2-1)\nonumber\\
	& \times \frac{g_q((1-\pi)^{k_1-1},(1-\pi)^{k_2-1})^2\langle s_2\rangle}
	{A(k_1,k_2,\pi)},
\end{align}
where $g_D(x)$ is the generating function of the total vertex degrees.   

\section{Approximating $\pi_c$ for mixed clique sizes}
The average number of $k_i$ vertices reached from a $k_j$-clique vertex for $i,j\in\{1,2\}$, equals
\begin{equation}
	M_{k_i,k_j}=\frac{\langle s_is_j\rangle-\delta_{k_i,k_j}}{\langle s_j\rangle}\sum_{j=1}^{k_i-1}jh(k_i,j,\pi)
\end{equation}
where $\delta_{k_i,k_j}$ is the Kronecker delta. Thus, the matrix $M$ is a branching matrix that describes the average number of vertices of type $k_i$ attached to a randomly chosen clique-edge of type $k_j$. The average number of vertices at generation $j$ of the offspring distribution can be expressed in terms of $M^j$. Therefore, if the largest eigenvalue of $M$ becomes larger than one, a giant component forms~\cite{karrer2010}.

Again, we approximate the solution by a second-order polynomial in $\pi$. Therefore, similarly to the analysis in Section~\ref{sec:piccliques}, we only keep the terms $h(k_i,1,\pi)$ and $h(k_i,2,\pi)$. Then, the condition on the largest eigenvalue of $M$ becomes~\cite{karrer2010}
\begin{align}
    & E_{k_1,k_1}(\pi+\pi^2(k_1-2))+E_{k_2,k_2}(\pi+\pi^2(k_2-2))\nonumber\\
    & = (\pi+\pi^2(k_1-2))(\pi+\pi^2(k_2-2))\nonumber\\
    & \quad \times \left(E_{k_1,k_1} E_{k_2,k_2}- E_{k_1,k_2} E_{k_2,k_1}\right)+1
\end{align}
where 
\begin{equation}
  E_{k_i,k_j}=\left(\frac{\langle s_is_j\rangle}{\langle s_j\rangle}-\delta_{k_i,k_j}\right)(k_i-1).
\end{equation}
Keeping only second order terms in $\pi$ yields
\begin{align}
    &  \pi^2\big(E_{k_1,k_1} E_{k_2,k_2}- E_{k_1,k_2} E_{k_2,k_1}-E_{k_1,k_1}(k_1-2)\nonumber\\
     & \quad -E_{k_2,k_2}(k_2-2)\big)-\pi\left(E_{k_1,k_1}+E_{k_2,k_2}\right)1=0
\end{align}

This equation has its positive solution as
\begin{widetext}
\begin{equation}
	\pi_c=\frac{-E_{k_1,k_1}-E_{k_2,k_2}+\sqrt{(E_{k_1,k_1}+E_{k_2,k_2})^2-4\left(E_{k_1,k_1} E_{k_2,k_2}- E_{k_1,k_2} E_{k_2,k_1}-E_{k_1,k_1}(k_1-2)-E_{k_2,k_2}(k_2-2)\right)}}{2\left(E_{k_1,k_1} E_{k_2,k_2}- E_{k_1,k_2} E_{k_2,k_1}-E_{k_1,k_1}(k_1-2)-E_{k_2,k_2}(k_2-2)\right)}.
\end{equation}
\end{widetext}

\section{Equality of numerator second term}\label{sec:eqsecondterm}
We now show that the numerator of the second approximating term in~\eqref{eq:Sgeneral} only depends on the degree distribution of the random graph, but not on its clique structures.
\begin{align}
	g_p((1-\pi)^{k-1})& =\frac{1}{\langle s \rangle}\sum_i ip_i(1-\pi)^{(k-1)(i-1)}\nonumber\\
	& =\frac{1}{\langle d \rangle}\sum_i i(k-1)p_i(1-\pi)^{i(k-1)-(k-2)}\nonumber\\
	&= g_{D^*-1}(1-\pi)(1-\pi)^{-(k-2)},
\end{align}
where $D^*$ is the size-biased degree distribution. Thus,
\begin{align}
	& \langle s \rangle g_p((1-\pi)^{k-1})^2(k-1)(1-\pi)^{2(k-2)}\pi\nonumber\\
	& =\langle d \rangle \pi  g_{D^*-1}(1-\pi),
\end{align}
which is independent of the clique size $k$, and only depends on the degree distribution.

\end{document}